\documentclass[aps,twocolumn]{revtex4-1}
\usepackage{graphicx}
\usepackage[justification=justified,width=\linewidth]{caption}
\usepackage{subcaption}
\usepackage{amsmath}
\usepackage{amsfonts}
\usepackage{amsthm}
\usepackage{amssymb}
\usepackage{amsbsy}
\usepackage{wasysym}
\usepackage{bm}
\usepackage{mathrsfs}
\usepackage{color}
\usepackage{times}
\usepackage[resetlabels]{multibib}

\begin{document}
\title{Characteristics and Correlation of Nonaffine Particle Displacements in the Plastic Deformation of Athermal Amorphous Materials}
\author {Meenakshi L}
\author {Bhaskar Sen Gupta}
\email{bhaskar.sengupta@vit.ac.in}
\affiliation{Department of Physics, School of Advanced Sciences, Vellore Institute of Technology, Vellore, Tamil Nadu - 632014, India}
\date{\today}

\begin{abstract} 
When an amorphous solid is deformed homogeneously, the response exhibits heterogeneous plastic instabilities with localized cooperative rearrangement of cluster of particles. The heterogeneous behavior plays an important role in deciding the mechanical properties of amorphous solids. In this paper, we employ computer simulation to study the characteristics and the spatial correlations of these clusters characterized by the non-affine displacements in amorphous solids under simple shear deformation in the athermal quasistatic limit. The clusters with large displacements are found to be homogeneously distributed in space in the elastic regime, followed by a localization within a system-spanning shear band after yielding. The distributions of the displacement field exhibit power-law nature with exponents strongly dependent on the deformation. The non-affine displacements show strong spatial correlations which become long-ranged with increasing strain. From our results, it is evident that the decay of the correlation functions is exponential in nature in the elastic regime. The yielding transition is marked by an abrupt change in the decay after which it is well described by power-law. These results demonstrate a scale-free character of non-affine correlations in the stead flow regime. These results are found to be robust and independent of the strain window over which the total non-affine displacement is calculated.

\end{abstract}
\maketitle

\section{Introduction} 
There has been a long-standing interest to understand the mechanical instabilities of amorphous materials because of their paramount importance in a wide range of fields, from fracture of metallic glasses in material engineering \cite{metglassfrac1,metglassfrac2} to earthquake and landslide in geology \cite{earthquake}. Despite extensive research, the physics of mechanical failure in amorphous materials is still poorly understood compared to its crystalline counterparts. Unlike crystalline solids where the topological defects or dislocations are considered as the basis of deformation and plastic flow \cite{crystal}, there is no long-range order in amorphous materials compared to which defects can be defined. Therefore, understanding amorphous plasticity requires an entirely different approach from crystalline plasticity.

The elementary process of amorphous plasticity involves the shear localization under the influence of mechanical loading and the subsequent collective rearrangement of a cluster of particles, often referred to as the shear transformation zone (STZ) \cite{argon,falk}.  Rearrangements are found to occur at any strain ($\gamma$) value. For small $\gamma$, an amorphous sample shows elastic behavior, and the stress ($\sigma$) increases linearly with $\gamma$. However, this branch is punctuated by localized plastic instabilities, where the irreversible structural rearrangements occur locally \cite{argon1,schall}. The typical size of such STZs is extended up to a few particle diameter. With increasing $\gamma$, these plastic drops trigger more events nearby and this correlated process spreads across the material in an avalanche-like manner, and subsequently the material yields via system spanning shear band formation \cite{Alexander, Bonn, 04VBB, 04ML, 05DA, 06TLB, 06ML, 09LP, 11RTV, 06SLG, 13KTG, 13NSSMM}.

Elucidating the influence on a plastic event by its predecessors spatially would be a significant advance in understanding the unwanted mechanical failure. Numerous experimental and computer simulation studies reported the existence of correlations between the STZs. For example, using confocal microscope in a shear colloidal glass it was observed that the global, as well as local non-affine displacement fields show strong spatial correlations in the plastic flow region \cite{chikkadi}. The long-range correlations that exhibit power-law decay originate from the elastic deformation field due to plastic events. These findings were substantiated by numerical simulations on hard-sphere glasses \cite{mandal,varnik}. In a similar spirit, spatial autocorrelation in the non-affine deformation field of sheared bulk metallic glasses was studied using computer simulations \cite{murali}. The decay of the correlation function was found to be exponential in nature. This was related to a length scale associated with the size of the STZs in the pre-yield regime. In ref. \cite{cubuk} results of experiments and simulations on a wide range of amorphous materials were reported and the correlation was found to be decaying exponentially. Finite rate deformation of Lennard Jones glasses at nonzero temperature showed a transition from exponential to power-law decay of correlation functions across the yielding transition \cite{nikolai}. Therefore, it is evident that the long debate over power-law vs exponential decay in the correlations of non-affine displacement fields is not settled yet. 

The main objective of our work is to gain insight into the characteristics of the non-affine displacement field and the nature of their spatial correlations in the elastic and steady-state plastic region. We, therefore, focus on the microscopic analysis of the non-affine displacement field of particles in a simple amorphous solid under athermal quasistatic shear (AQS) using computer simulation. The principle utility of the AQS algorithm is that it enables us to probe the shear-induced structural changes of the system in the absence of thermal fluctuations \cite{Lacks1999_JCP, Barrat2002_PRB, Lemaitre2004_PRL1, Lemaitre2004_PRL2, Lemaitre2006_PRE, Procaccia2009_PRE}. Our results show that in the elastic branch, the STZs are homogeneously distributed in space across the system followed by the system spanning shear band formation post yielding. This is reflected in the distinct change of the decaying nature of the correlation function of the non-affine displacement. The spatial extent of the displacement field is examined by computing the distribution function which shows power-law decay, with an exponent dependent on the deformation. Finally, we revisit the riddle of exponential vs power-law decay of the correlations between STZs in the pre and post-yield regime. The robustness of these results is tested by computing the non-affine displacements as a function of accumulated deformation as well as various strain windows.

Our paper is structured as follows: a detailed description of the sample preparation using molecular dynamics simulations and the athermal quasistatic shear protocol is given in the next section. In section III we discuss the methods to calculate the nonaffine displacement field and present results and analysis of their characteristics and spatial correlation functions. Finally, a conclusion is presented in Sec. IV. 
\section{Numerical Simulation Details}
To prepare amorphous samples we resort to molecular dynamics (MD) simulations using the well studied Kob-Anderson binary mixture model \cite{Kob}. The system comprises of two types of particles labeled as $A$ and $B$ with their number ration $80:20$. We simulate $N=32400$ particles confined in a cubic box of volume $V$. For simplicity, the mass $m$ of both types of particles is taken to be the same and equal to unity. The particles interact via a pairwise Lennard-Jones potential given by
\begin{eqnarray}
U_{\alpha\beta}(r) &=& 4\epsilon_{\alpha\beta}\Big[\Big(\frac{\sigma_{\alpha\beta}}{r}\Big)^{12} - \Big(\frac{\sigma_{\alpha\beta}}{r}\Big)^{6}\Big], r\le r_{cut} \nonumber \\
&=& 0 \,~~~~~~~~~~~~~~~~~~~~~~~~~~~~~~~~ r\textgreater r_{cut}, \label{Uij}
\end{eqnarray}
where $\alpha, \beta \in  \rm{A, B}$. The units of various quantities in our simulation are as follows: length is expressed in the unit of $\sigma_{AA}$, energy in the unit of $\epsilon_{AA}$, time in the unit of $(m\sigma_{AA}^2/48\epsilon_{AA})^{1/2}$ and temperature in the unit of $\epsilon_{AA}/k_{\rm B}$. Here $k_{\rm B}$ is the Boltzmann constant which is unity. The parameters $\sigma_{\alpha\beta}$ and $\epsilon_{\alpha\beta}$ are chosen as follows: $\sigma_{AA}=1.0, \sigma_{BB}=0.8, \sigma_{AB}=0.88$ and $\epsilon_{AA}=1.0, \epsilon_{BB}=0.5, \epsilon_{AB}=1.5$. The asymmetric choice of the parameters introduces frustration in to the system which inhibits crystallization below $T_g$ and forms stable glass. The cutoff $r_{cut}=2.5$ is used to reduce computational efforts. 

MD simulations are employed by integrating Newton's equation of motion using velocity Verlet algorithm \cite{Verlet} with a discretization time step $\Delta t=0.005$ in Lennard-Jones units. We work in the $NVT$ ensemble at a constant density $\rho=N/V=1.2$ with periodic boundary conditions in all directions. The temperature of the system is controlled by the Berendsen thermostat \cite{Berendsen}. To prepare the glass samples, the binary mixture is first equilibrated in the liquid state at temperature $T=1.0$. The glass transition temperature for the KA model at this density is $T_g=0.435$ \cite{Kob}. After equilibration, the system is cooled linearly with a rate $10^{-5}$ to a very low temperature $T=0.001$ which is sufficiently low to eliminate any appreciable thermal effects. Finally, the system is brought to the minimum energy state using conjugate gradient energy minimization algorithm where the temperature is formally $T = 0$. For statistical averaging 100 independent glassy samples are generated by repeating the whole process with different initial realizations.

To explore plastic failure the glassy samples are subjected to simple shear deformation using AQS protocol with the limit $T\rightarrow0$ and $\dot{\gamma}\rightarrow0$, where $\dot{\gamma}$ is the strain rate \cite{06ML}. The AQS algorithm includes two iterating steps: (a) a freshly quenched glassy sample is deformed by applying an affine simple shear transformation to each particle $i$ of the system as
\begin{equation}
r_{ix} \rightarrow r_{ix} +r_{iz}\delta\gamma, ~~~r_{iy} \rightarrow r_{iy}, ~~~r_{iz} \rightarrow r_{iz}
\end{equation}
using the Lees-Edwards boundary conditions \cite{lees}. We choose sufficiently small strain increment $\delta\gamma=10^{-4}$. (b) After every affine transformation step, the potential energy of the deformed system is minimized using conjugate gradient algorithm under the constraints imposed by the boundary condition. By repeating steps (a) and (b) we can reach up to arbitrarily large strain values. The AQS method ensures that the system is in mechanical equilibrium after every differential strain increase and follows strain-induced changes of the potential energy surface. Therefore, the change of particles position does not map homogeneously with the macroscopic strain but evolves as $\textbf{r}_i=\textbf{r}_i'+\textbf{u}_i$, where $\textbf{u}_i$ represents the nonaffine displacement of the $i$th particle. Under the influence of increasing strain, the energy minimum changes its shape continuously until it develops zero-curvature along a certain direction. Any further infinitesimal strain increment causes the system to fall to a new minimum with a large non-affine displacement of particles, identified as plastic event. During AQS deformation the particle configurations are saved at regular intervals of strain values for further analysis. 
\section{Results}
\begin{figure}[!ht]
	\centering
	\includegraphics[width=\columnwidth]{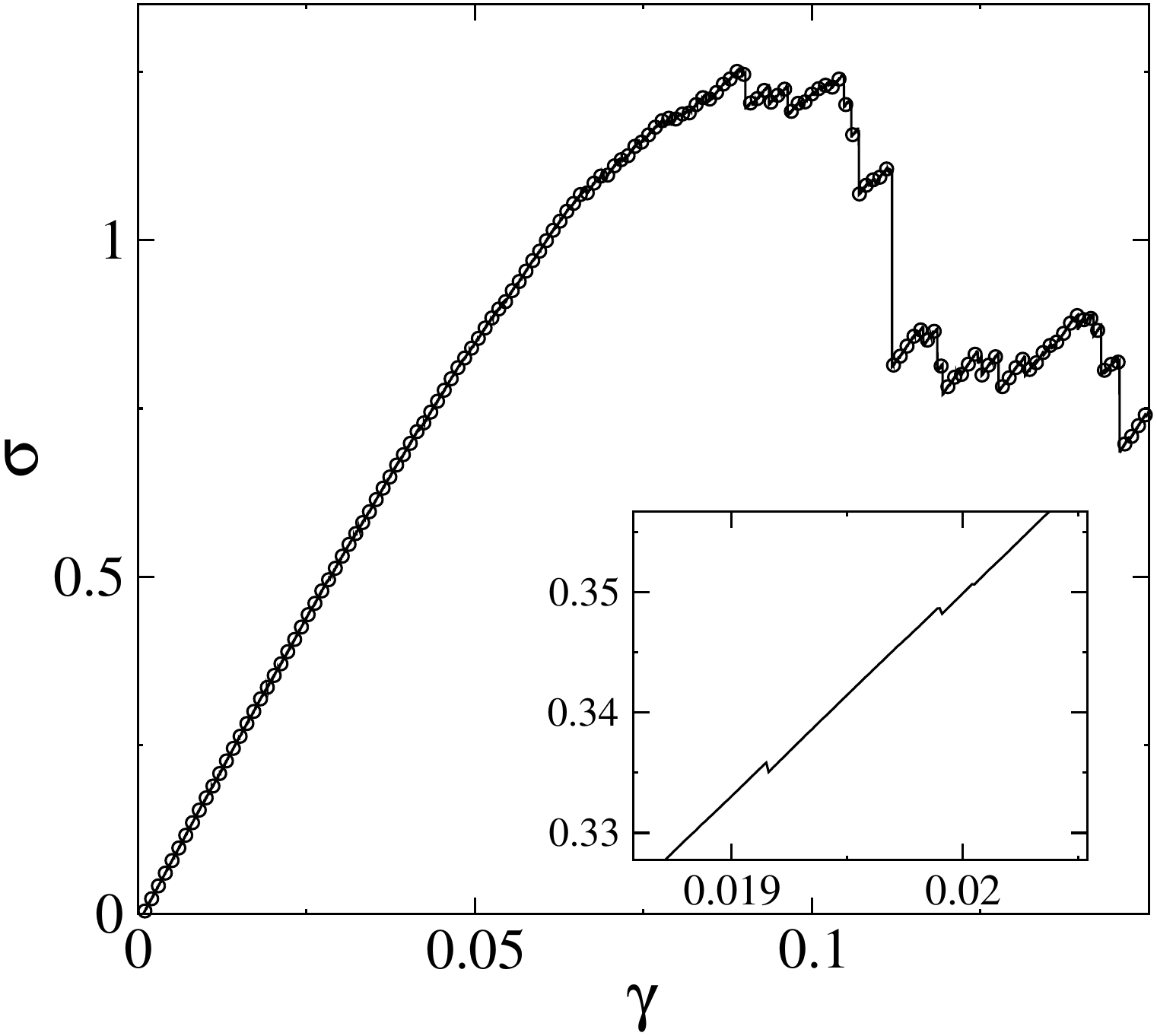}
	\caption{Typical stress $(\sigma)$ vs strain $(\gamma)$ curve for an individual glassy sample of $N=32400$ particles under athermal quasistatic simple shear deformation. Inset shows the same curve zoomed over a small strain window in the elastic regime.}
	\label{fig1}
\end{figure}
In Fig. 1 we show the typical stress vs strain curve obtained from the AQS simulation. The system yields at around $\gamma \approx 8\%$ and after that steady-state plastic flow sets in. The mechanical response comprises a series of plastic events involving the rearrangement of cluster of particles that produces long-range displacement fields. To investigate the heterogeneous response, we resort to the microscopic analysis involving spatial configurations of atoms with large relative non-affine displacements associated with the deformation events. More precisely, we compute the non-affine displacements of the local shear transformations in deformed solids originally proposed by Falk and Langer \cite{falk} with some important modifications involving the strain $\gamma$, defined as
\begin{equation}
\begin{split}
D^2(\gamma,\Delta \gamma)  = & \frac{1}{N_i} \sum\limits_{j=1}^{N_i}  \lbrace \boldsymbol{\mathrm{r}}_j(\gamma) - \boldsymbol{\mathrm{r}}_i(\gamma) - \\
&   \boldsymbol{\mathrm{J}}_i [\boldsymbol{\mathrm{r}}_j(\gamma-\Delta \gamma) - \boldsymbol{\mathrm{r}}_i(\gamma-\Delta \gamma)]\rbrace^2
\end{split}
\label{eq-d2}
\end{equation}
Here $\boldsymbol{\mathrm{r}}_i(\gamma)$ is the position vector of the $i$th particle at the strain value $\gamma$, $\boldsymbol{\mathrm{J}}_i$ is the transformation matrix that maps the $i$th particle and its first nearest neighbor at strain $\gamma$ and $\gamma-\Delta \gamma$ respectively via an affine deformation. This quantity proposed in ref. \cite{falk} is extensively used for the spatio-temporal analysis of non-affine displacements in shear-driven amorphous solids \cite{varnik,cubuk,jana,chikkadi,ding,Nikolai1,Nikolai2,vijay1}. The choice of $\Delta \gamma$ in Eq. \ref{eq-d2} plays an important role in the resolution of different plastic events and the nature of $D^2$ field. In this work, we present results for two different choices of the reference frame with respect to which the $D^2$ field is computed; (a) the undeformed atomic configuration which corresponds to $\Delta \gamma=\gamma$. In this case, the accumulation of all the plastic events that occurred during the total deformation $\gamma$ will contribute to the $D^2$ field calculation. (b) The reference frame is at a fixed distance $\Delta \gamma$ from the applied strain $\gamma$ and thus maintains a constant strain window. Here the plastic events occurring within the deformation interval $\gamma-\Delta \gamma$ to $\gamma$ are considered. 

\begin{figure}[h]
	\centering
	\begin{subfigure}[h!]{0.4\columnwidth}
		\includegraphics[width=\columnwidth]{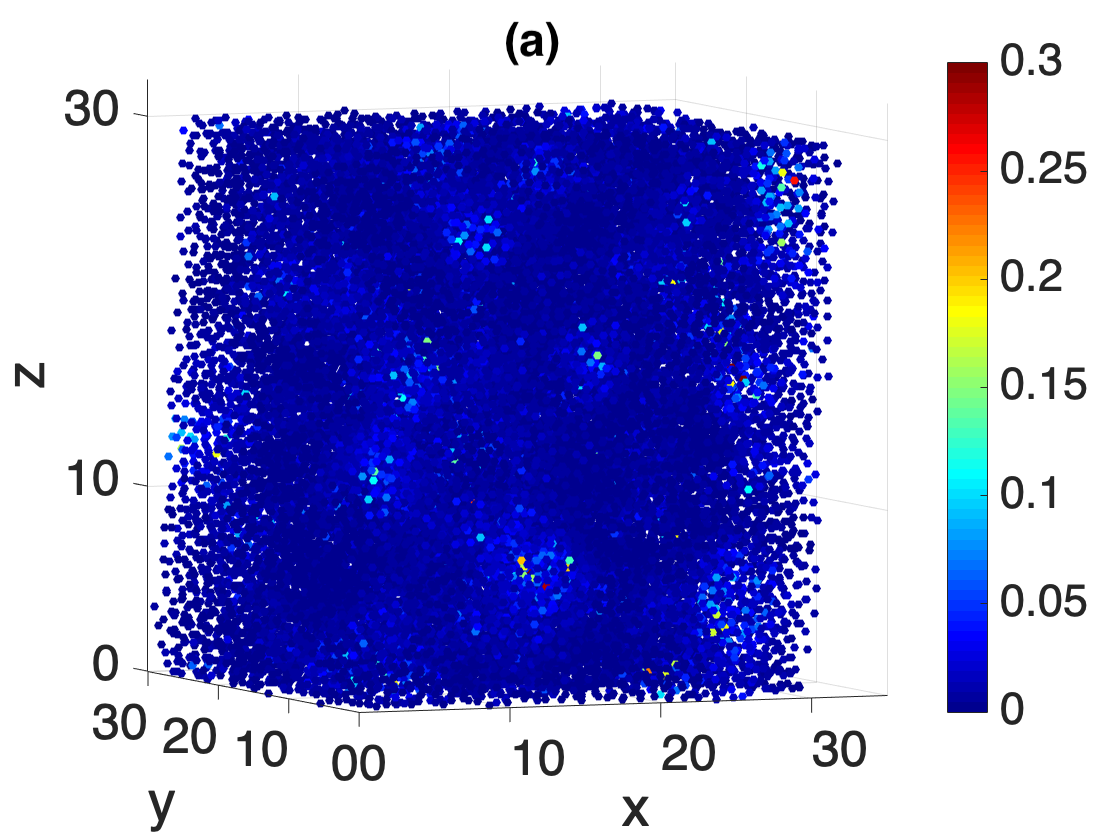}
	\end{subfigure}%
	\begin{subfigure}[h!]{0.4\columnwidth}
		\centering
		\includegraphics[width=\columnwidth]{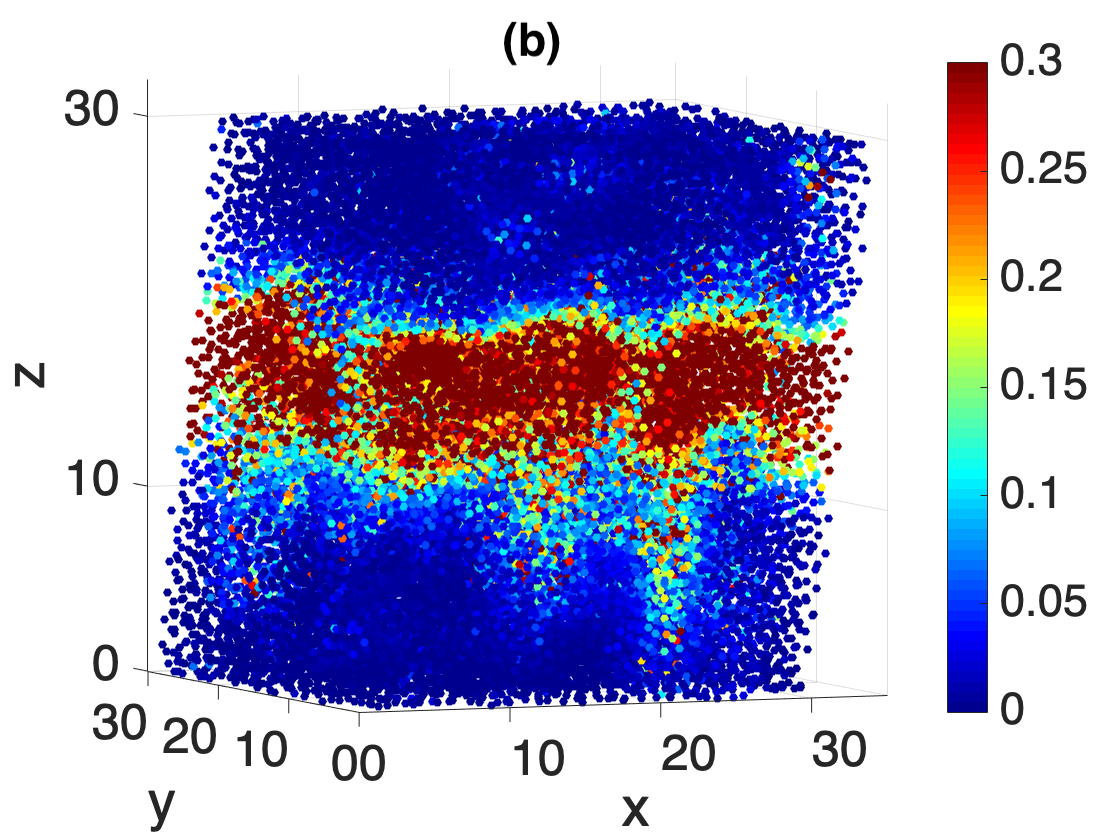}
	\end{subfigure}\\
\begin{subfigure}[h!]{0.4\columnwidth}
	\includegraphics[width=\columnwidth]{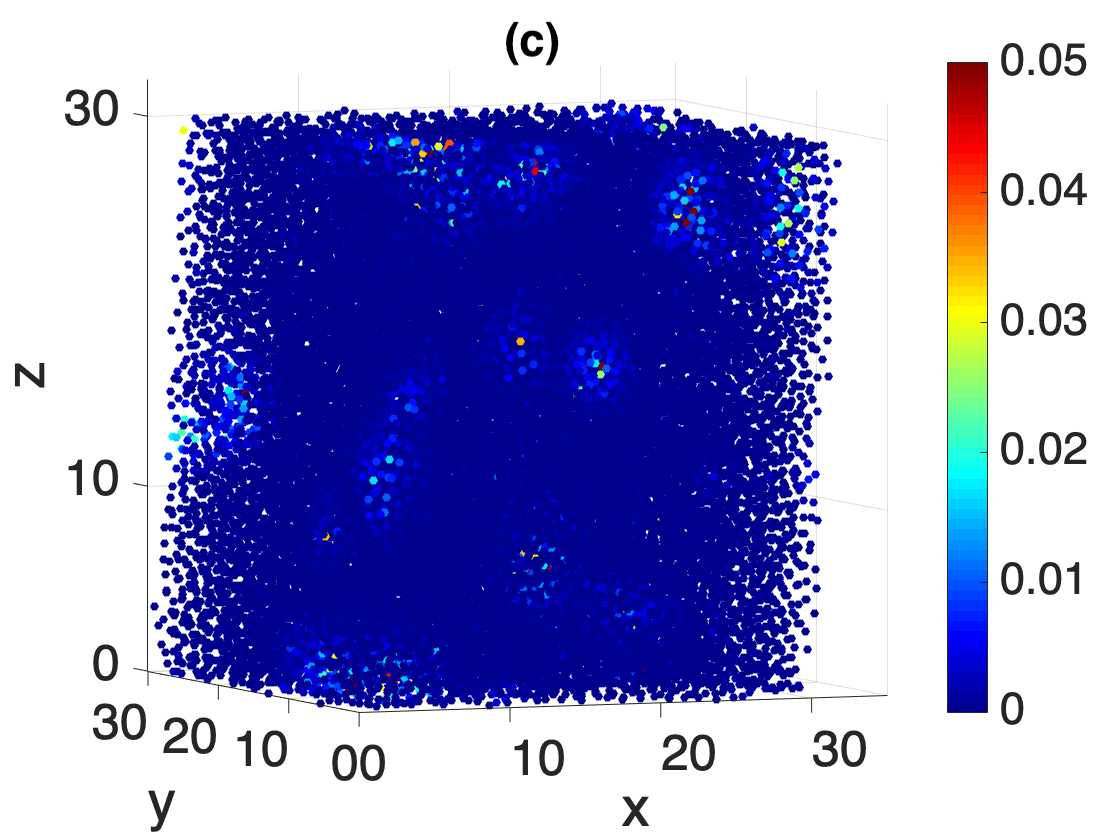}
\end{subfigure}%
\begin{subfigure}[h!]{0.4\columnwidth}
	\centering
	\includegraphics[width=\columnwidth]{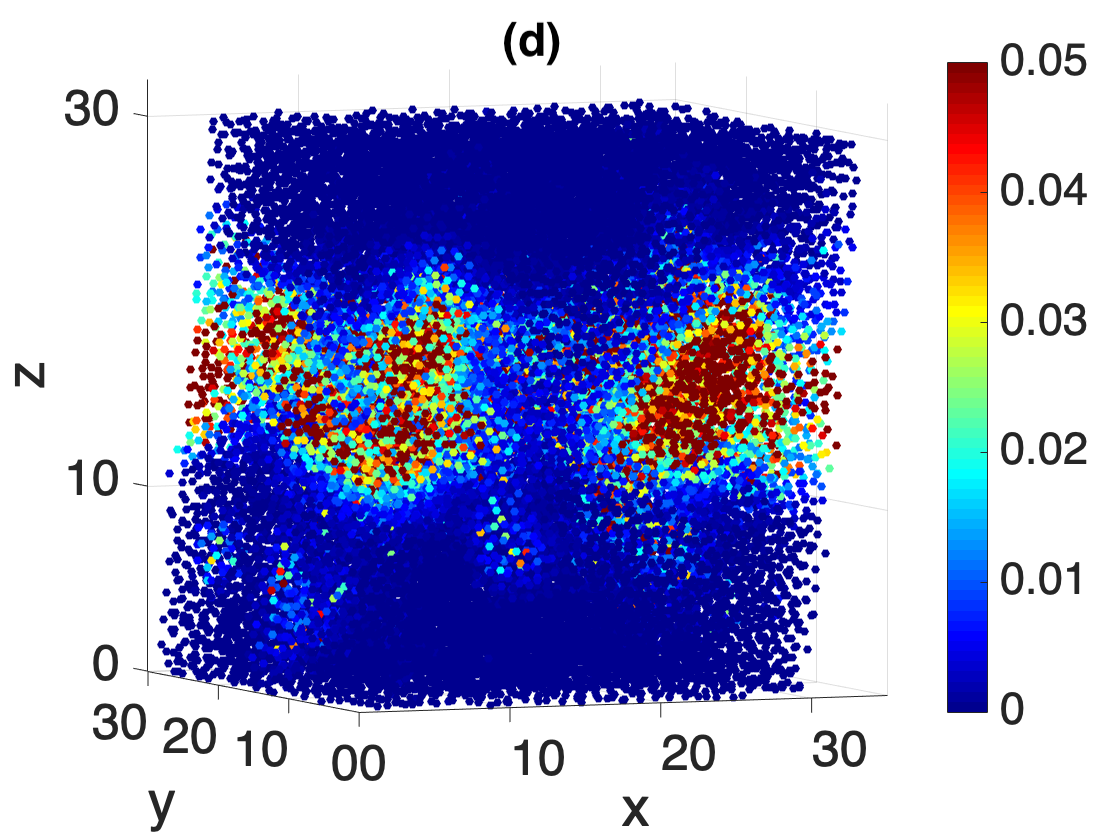}
\end{subfigure}
	\caption{Typical snapshots of the deformed glassy samples for (a) $\Delta \gamma=\gamma$ at $\gamma=0.07$, (b) $\Delta \gamma=\gamma$ at $\gamma=0.12$, (c) $\Delta \gamma=1\%$ at $\gamma=0.07$ (d) $\Delta \gamma=1\%$ at $\gamma=0.12$. The particles are color-coded according to their non-affine displacements $D^2(\gamma,\Delta \gamma)$ (see text).}
	\label{snap}
\end{figure}

The representative snapshots for the deformed glassy states in the elastic and steady state regime are shown in Fig. \ref{snap}. We compute the displacement field $D^2(\gamma, \Delta\gamma)$ for these configurations from Eq. \ref{eq-d2} for two  different choices of reference frames:  $\Delta \gamma=\gamma$ and $\Delta \gamma=1\%$.
\onecolumngrid\
\twocolumngrid\

In Fig. \ref{snap} we color-code the particles according to their non-affine displacement field. It can be clearly seen that the STZs are homogeneously distributed in space in the elastic regime. The scenario changes completely as the mobile particles localize in a band-type structure running across the system and form shear band.
\begin{figure}
	\centering
	\begin{subfigure}[h!]{0.45\columnwidth}
		\includegraphics[width=\columnwidth]{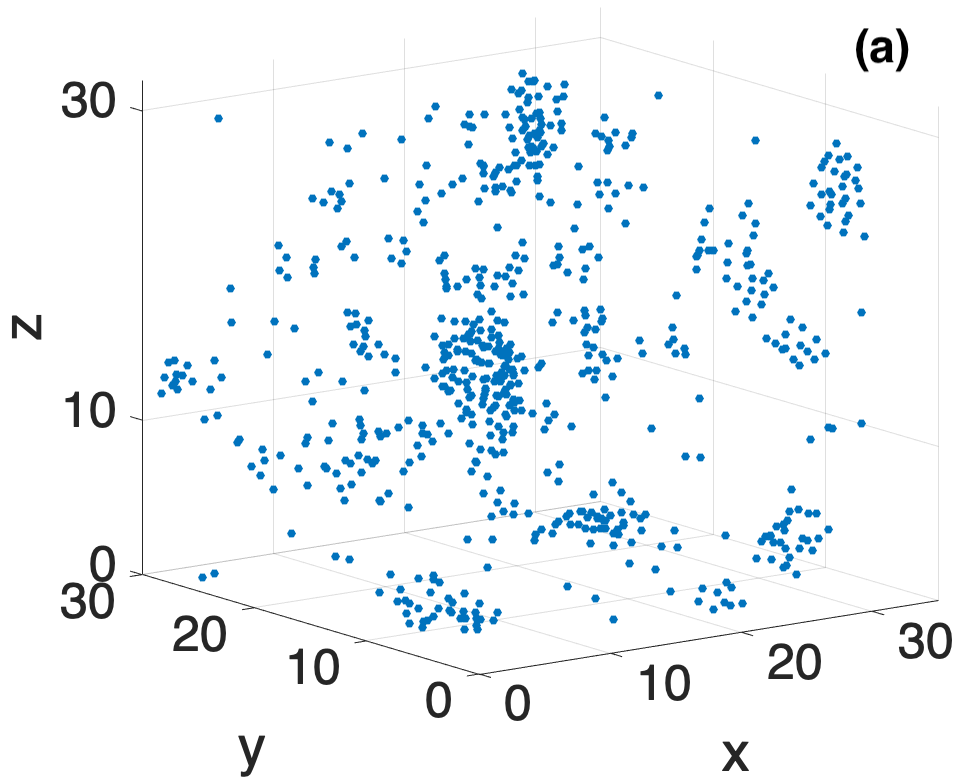}
	\end{subfigure}%
	\begin{subfigure}[h!]{0.45\columnwidth}
		\centering
		\includegraphics[width=\columnwidth]{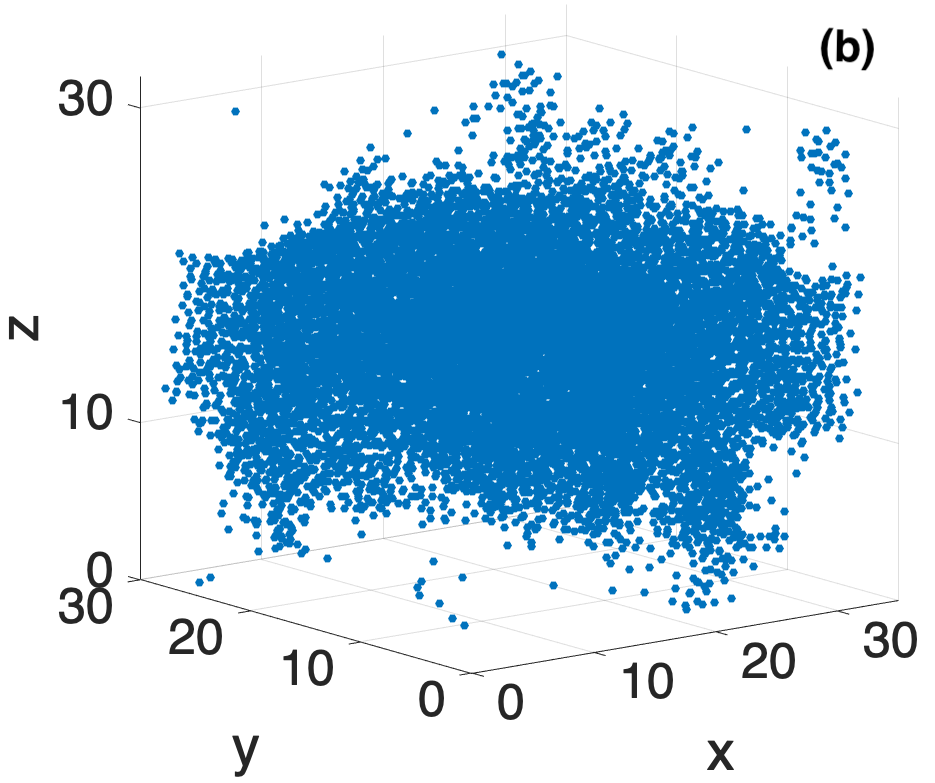}
	\end{subfigure}\\
	\begin{subfigure}[h!]{0.45\columnwidth}
		\includegraphics[width=\columnwidth]{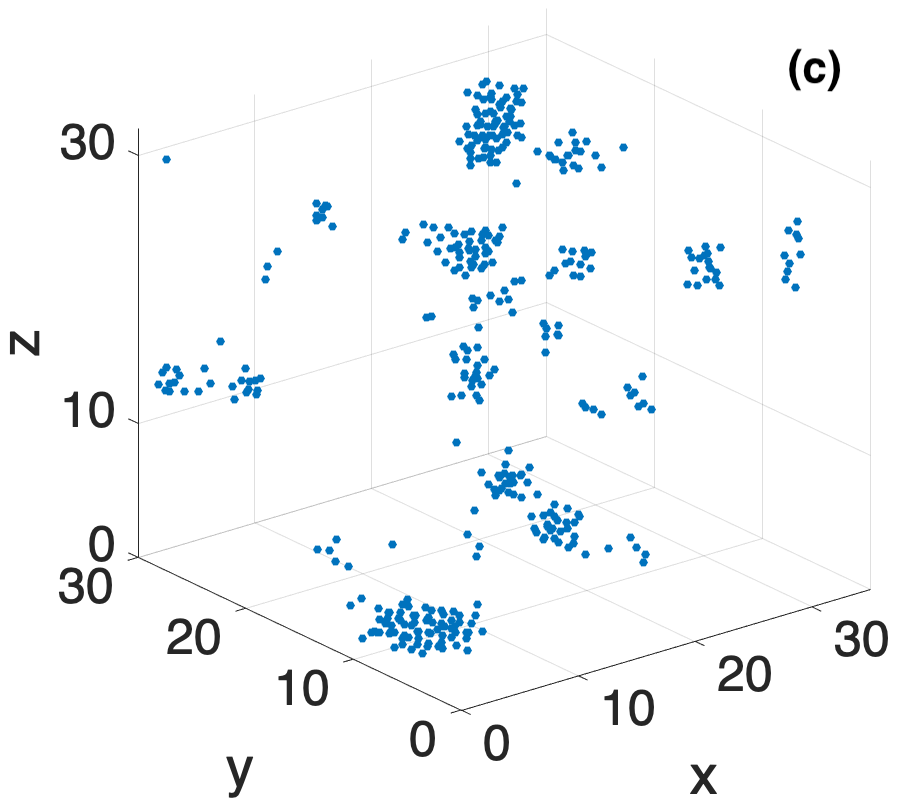}
	\end{subfigure}%
	\begin{subfigure}[h!]{0.45\columnwidth}
		\centering
		\includegraphics[width=\columnwidth]{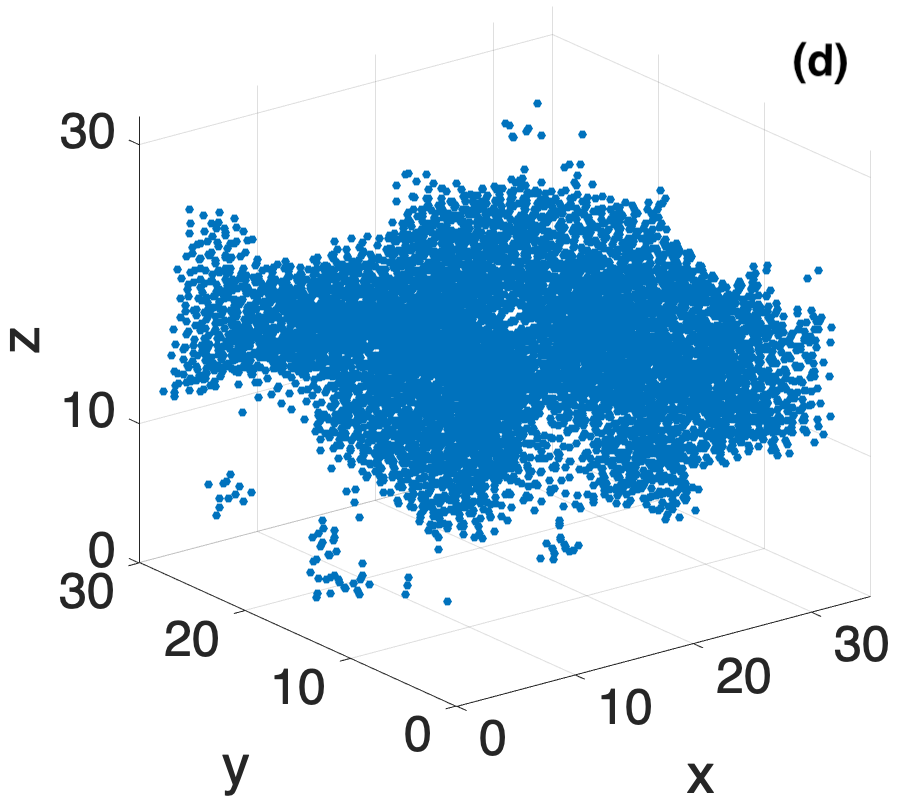}
	\end{subfigure}
	\caption{Spatial configurations of particles with large non-affine displacements in Fig. \ref{snap} for (a) $D^2>0.05$, (b) $D^2>0.05$, (c) $D^2>0.01$ and (d) $D^2>0.01$ }
	\label{snapcluste}
\end{figure}

The spatial distribution of clusters is further examined by plotting the configurations of particles with large nonaffine displacements. This is shown in Fig. \ref{snapcluste}. In the elastic regime, we observe the aggregation of particles into disconnected clusters that are uniformly distributed in the sample. Post yielding the clusters tend to merge and form shear band of the order of system size.

\begin{figure}
	\centering
	\includegraphics[width=0.9\columnwidth]{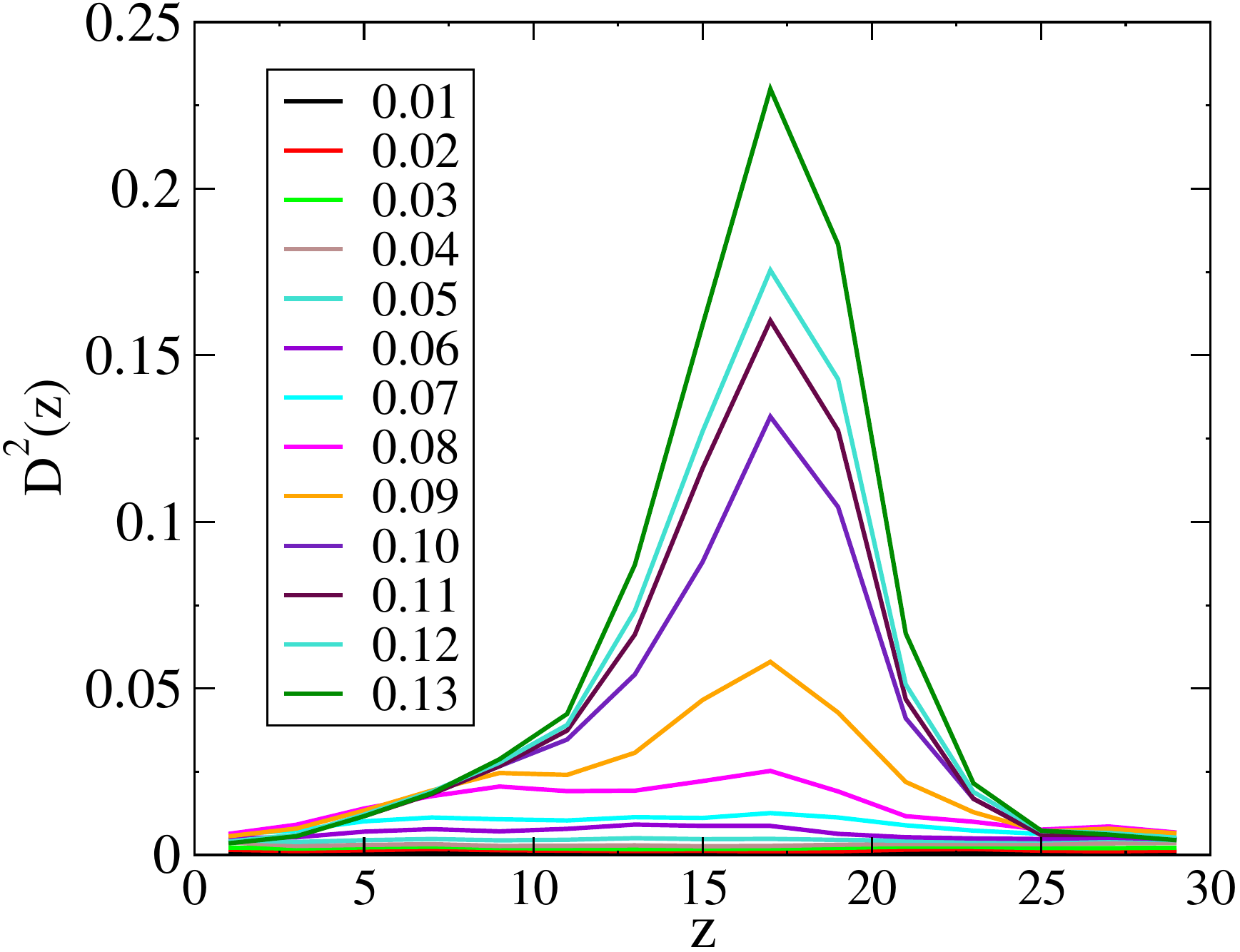}
	\caption{The spatial distribution of the non-affine displacements $D^2(z)$ vs $z$ for $\Delta \gamma=\gamma$ at different strain values mentioned in the inset. The $D^2(z)$ shown here corresponds to the same sample in Fig. \ref{snap}.}
	\label{D2z}
\end{figure}

This can be better understood by the probability distribution of the displacement field which involves spatially averaged profiles of $D^2(z)$ computed over a small bin of thickness $\Delta z$. This is shown in Fig. \ref{D2z} for a chosen set of strain values with the undeformed state as the reference frame. 

For small deformation, the distribution curves are almost flat, indicating that the clusters of mobile particles are homogeneously distributed across the sample. The average value of $D^2(z)$ increases gradually with strain. The yielding transition is marked by the appearance of a distinct peak in the distribution, indicating the formation of shear band. The typical width of the shear band is of several particle diameters. The width of the peak increases gradually with strain implying the expansion of the shear band. 

\begin{figure}
	\centering
	\begin{subfigure}[h!]{1.0\columnwidth}
		\includegraphics[width=0.8\columnwidth]{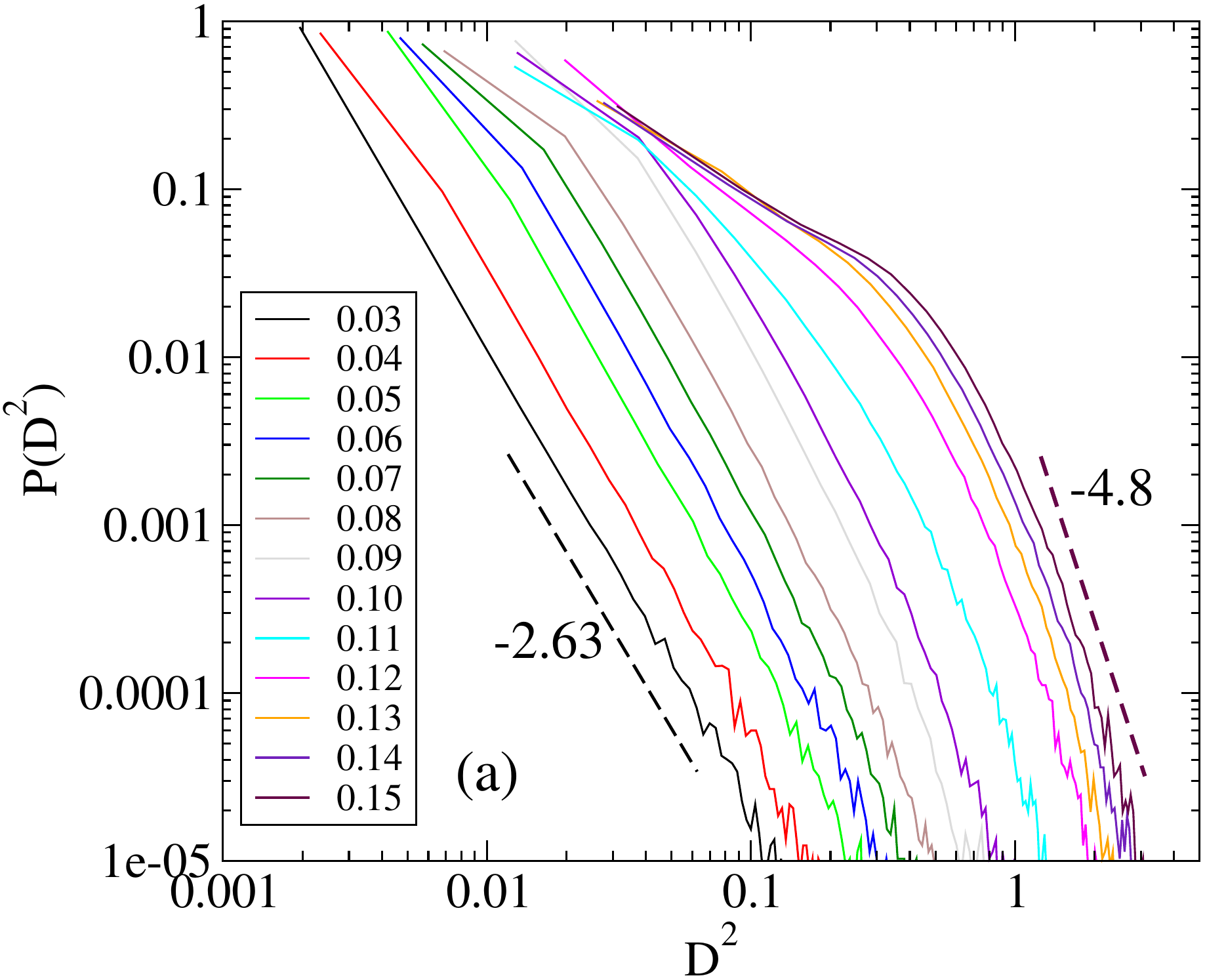}
	\end{subfigure}\\
	\begin{subfigure}[h!]{1.0\columnwidth}
		\centering
		\includegraphics[width=0.8\columnwidth]{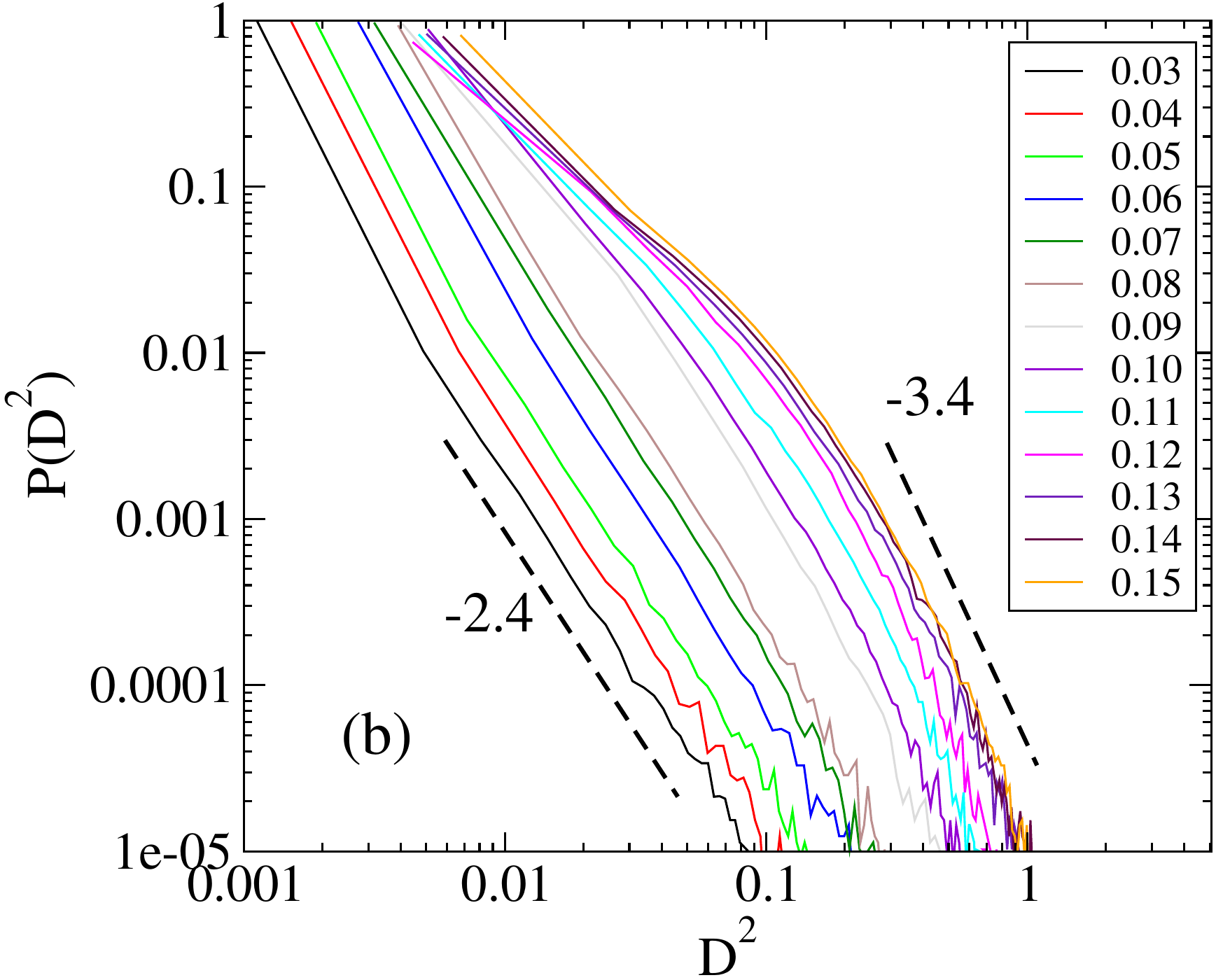}
	\end{subfigure}
	\caption{The probability distribution of the non-affine displacements for (a) $\Delta \gamma=\gamma$ and (b) $\Delta \gamma=1\%$ for the strain values mentioned in the inset. The dashed lines indicate the slope.}
	\label{PDF}
\end{figure}

To understand the role of $\Delta \gamma$ on the nature of the nonaffine displacements and their spatial extent, we compute the probability distribution functions (PDFs) of $D^2(\gamma, \Delta \gamma)$ for the two different choices of $\Delta \gamma$ mentioned above for a set of strain values below and above yielding. This is shown in Fig. \ref{PDF}. With increasing strain amplitude, the tail of the PDFs shifts towards larger $D^2(\gamma, \Delta \gamma)$ values. In both cases, the PDFs exhibit a power-law decay for all strain values. However the power-law exponent changes with strain. In the case of undeformed reference frame ($\Delta \gamma=\gamma$), the exponent varies from about -2.63 to about -4.8 for the range of deformations chosen in our simulation. Incidentally, our result near yielding is very close to the exponent -2.8  observed experimentally in the steady-state shear of colloidal glasses \cite{chikkadi}. When the second choice of reference frame $\Delta \gamma=1\%$ is used to compute $D^2(\gamma, \Delta \gamma)$, the exponent changes from about -2.4 to about -3.4 for the same set of strain values.  In both cases, the distribution functions become wider with strain in the elastic regime. But post yielding, unlike the first case, for $\Delta \gamma=1\%$, the PDFs start to converge above $\gamma=13\%$ with an exponent -3.4. After this, the exponent becomes insensitive to higher strain values. Therefore, we find the results are strongly dependent on the choice of reference frame for computing the nonaffine displacements. Hence, we can anticipate that the method to compute the $D^2(\gamma, \Delta \gamma)$ field will play a major role in determining the correlation between STZs.

\begin{figure}
	\centering
	\begin{subfigure}[h!]{0.8\columnwidth}
		\includegraphics[width=\columnwidth]{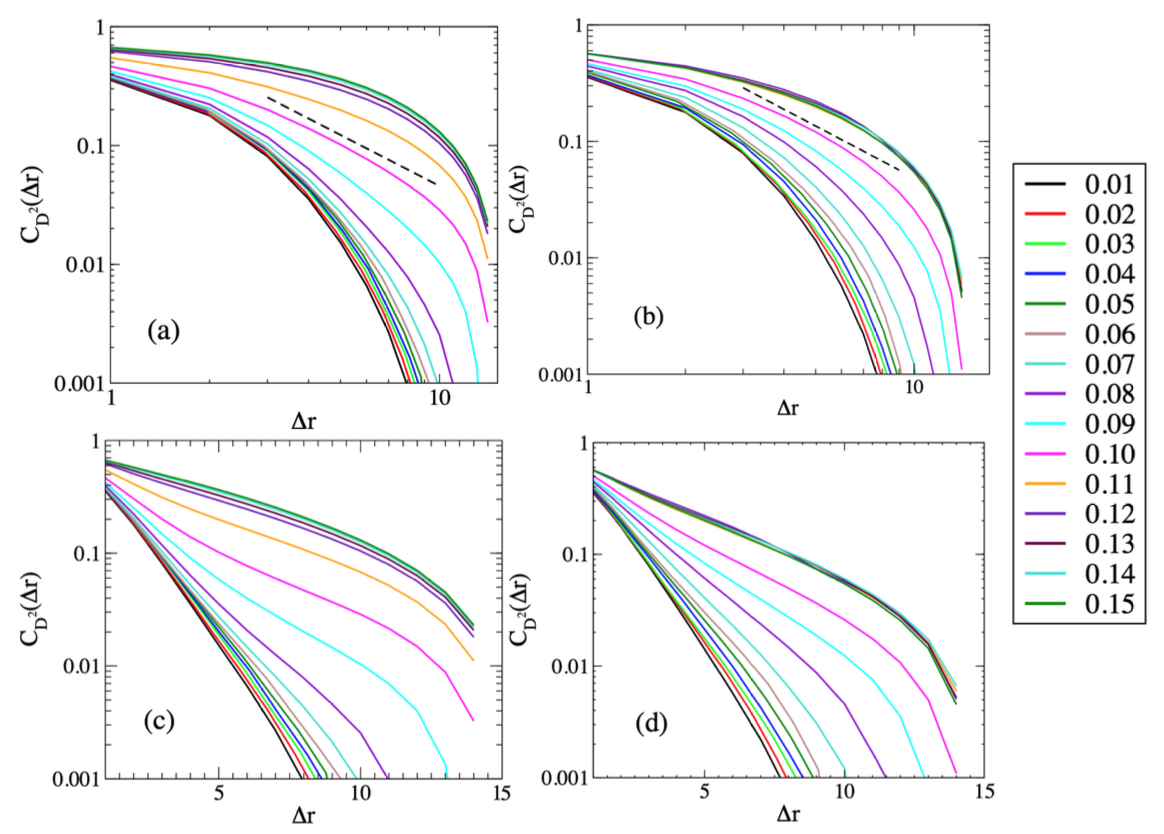}
	\end{subfigure}%
	\caption{Spatial correlation function of the non-affine displacement field for (a) $\Delta \gamma=\gamma$ and (b) $\Delta \gamma=1\%$ in the log-log scale for the strain values mentioned on the right. The dashed liens represent the slope -1.3 (see text). In figs. (c) and (d) we show the same data of (a) and (b) respectively in the log-linear scale.}
	\label{cor}
\end{figure}

To quantify the geometry of plastic events over space, we compute the spatial correlations of $D^2(\gamma, \Delta\gamma)$ field as follows: 
\begin{equation}
C_{D^2}(\Delta \textbf{r}) = \frac{\langle D^2(\textbf{r}+\Delta \textbf{r})  D^2(\textbf{r})\rangle-\langle D^2(\textbf{r})\rangle^2}{\langle D^2(\textbf{r})^2\rangle-\langle D^2(\textbf{r})\rangle^2}
\end{equation}
Here the angular brackets represent an average of spatial coordinate $\textbf{r}$ over 100 independent samples. In Fig. \ref{cor} we show the correlation results involving the nonaffine measure given by Eq. \ref{eq-d2} with respect to two different choices of reference frames mentioned above at different strain values.
\onecolumngrid\
\twocolumngrid\ 
As expected, in both cases, the correlation becomes long ranged with increasing strain. The relative 
change in $C_{D^2}(\Delta \textbf{r})$ abruptly becomes larger near the yielding transition ($\gamma 
\approx 8\%$) indicating the system has developed a system-spanning displacement field. This is 
consistent with the observation in the peak height in Fig. \ref{D2z}. 

Our results show the existence of long-range spatial correlation in the $D^2$ field. To characterize the nature of decay of $C_{D^2}(\Delta \textbf{r})$ we show in Fig. \ref{cor}  the correlation function in the log-linear scale for the choices $\Delta \gamma=\gamma$ and $\Delta \gamma=1\%$ at different strain values. In both cases, a constant slope in the semi-log plot is visible for $\gamma < 8\%$. These observations confirm the exponential decay of $C_{D^2}(\Delta \textbf{r})$ in the elastic regime and the existence of a characteristic length scale of plastic events. Post yielding the decay can be well described by a power-law decay. These results are found to be independent of the choice of the reference frame for $D^2(\gamma, \Delta \gamma)$ calculations. At large strain, the correlation functions gradually converge in the plastic flow regime. The convergence rate appears to be faster for $\Delta \gamma=\gamma$ than for cumulative strain. The power-law exponent close to the yielding transition is estimated to be -1.3. We mention here that the same exponent was reported in the experimental study of steadily sheared colloidal glasses \cite{vijay1}. This is shown in Fig. \ref{cor} with dashed lines for reference. The power-law decay demonstrates the scale-free character of the STZs in the stead flow regime. The deviation of the tail part can be attributed to the finite size of the system. We, therefore, conclude that the characteristic nature of the spatial correlations of non-affine displacements is entirely different in the elastic and the plastic regime and transforms from an exponential to a power-law decay at the yielding transition. These results are found to be robust and insensitive to the choice of $\Delta \gamma$.

\begin{figure}[!ht]
	\centering
	\includegraphics[width=0.9\columnwidth]{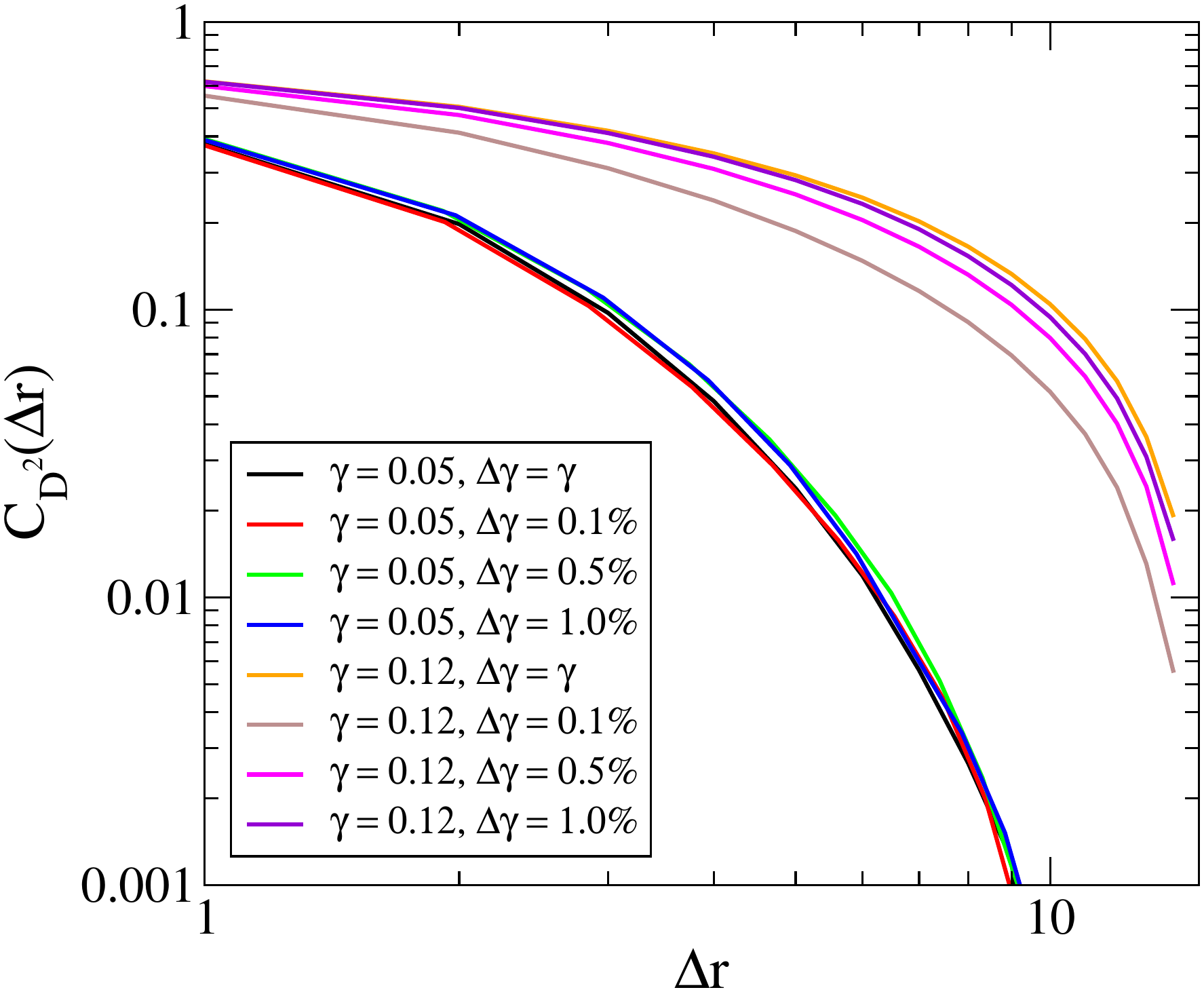}
	\caption{Spatial correlation function of the non-affine displacement field corresponding to the different choice of reference frame $\Delta \gamma=\gamma, 0.005, 0.01$ and $0.02$ at $\gamma=0.05$ and $0.12$}
	\label{corCompare}
\end{figure}

The role of the choice of reference frame on the correlation functions is further investigated. We compute $C_{D^2}(\Delta \textbf{r})$ with the undeformed configuration ($\Delta \gamma=\gamma$) as the reference frame as well as for three different strain windows $\Delta \gamma=0.1\%, 0.5\%$ and $1.0\%$. The results are shown in Fig. \ref{corCompare} for the elastic ($\gamma=5\%$) and steady state ($\gamma=12\%$) regime. 
We observe that before yielding the correlations overlap and therefore do not depend on the choice of the reference frame. Also, the nature of decay can be well described by exponential function over the entire range of distance. This behavior completely changes post yielding and we see a strong dependence on the reference frame. The correlation functions decay as power-law and become long ranged with increasing strain window.

\section{Summary and Conclusion}
In summary, we have performed computer simulations on a modeled amorphous solid to gain insight 
into the spatial distribution and structural connection of plasticity by computing the non-affine 
displacements of all the particles. The glassy samples were prepared by slow annihilation of the Kob 
Andersen binary LJ model to a very low temperature below $T_g$ and then simple shear was applied 
following AQS protocol. Different choices of reference frames ($\Delta \gamma$) to compute 
non-affine displacements were explored. The simulation results offer manifold conclusions. It was 
observed that, in the elastic regime, particles rearrangement with large non-affine displacements 
happens in isolated clusters that are relatively homogeneously distributed in space. Post-yielding it 
becomes inhomogeneous and localized by a system spanning shear band formation. The displacement 
field shows power-law distribution and the exponent strongly depends on the choice of $\Delta 
\gamma$. Our results for cumulative strain are found to be similar to sheared colloidal glass in the 
plastic flow regime.  Nonaffine displacements exhibit a strong spatial correlation in both the 
deformation regimes. One of the important results obtained from our work is the nature of the 
correlation function which changes from exponential to the power-law decay, and the yielding transition 
plays a pivotal role in this. Therefore, the long-range correlation function demonstrates a scale-free 
steady-state plastic regime. These results are found to be robust and independent of the choice of 
$\Delta \gamma$.
\newpage
\section*{Acknowledgment} 
B. Sen Gupta acknowledges Science and Engineering Research Board (SERB), Department of Science and 
Technology (DST), Government of India (no. SRG/2019/001923) for financial support. Meenakshi L. 
acknowledges VIT for doctoral fellowship. B. Sen Gupta would like to thank Vijaykumar Chikkadi for 
many useful discussions. \\


\section*{References} 
\begin{thebibliography}{}
\bibitem{metglassfrac1}
S. M. Chathoth, B. Damaschke, J. P. Embs, and K. Samwer, Appl. Phys. Lett. \textbf{95}, 191907 (2009).

\bibitem{metglassfrac2}
M. D. Demetriou, M. Floyd, C. Crewdson, J. P. Schramm, G. Garett, and W. L. Johnson, Scripta Materialia \textbf{65}, 799-802 (2011).

\bibitem{earthquake}
M. E. J. Newman, Contemp. Phys. \textbf{46}, 323 (2005); P. Bak, and C. Tang, J. Geophys. Res. \textbf{94}, 15635 (1989); Z. Olami, H. J. S. Feder, and K. Christensen, Phys. Rev. Lett. \textbf{68}, 1244 (1992).

\bibitem{crystal}
F. Seitz and D. Turnbull, Eds. Solid State Physics Vol 18 (Academic Press, New York, 1966, pp. 274-420; see also Vol. 19, 1966 pp. 1-134)	

\bibitem{argon}
A. S. Argon, Acta Metall. Mater. \textbf{27}, 47 (1979).

\bibitem{falk}
M.L. Falk and J.S. Langer, Phys. Rev. E \textbf{57}, 7192 (1998).

\bibitem{argon1}
A. Argon, H. Kuo, Mater. Sci. Eng. \textbf{39}, 101 (1979).

\bibitem{schall}
P. Schall, D.A. Weitz, F. Spaepen, Science \textbf{318}, 1895 (2007).

\bibitem{Alexander}
S. Alexander, Phys. Rep. \textbf{296}, 65 (1998).	

\bibitem{Bonn}
D. Bonn, and M. M. Denn, Science \textbf{324}, 1401 (2009).

\bibitem{04VBB}
F. Varnik, L. Bocquet, and J. L. Barrat, J. Chem. Phys. \textbf{120}, 2788 (2004).

\bibitem{04ML}
C. E. Maloney, and A. Lemaitre, Phys. Rev. Lett. \textbf{93}, 016001 (2004).

\bibitem{05DA}
M. J. Demkowicz, and A. S. Argon, Phys. Rev. B \textbf{72}, 245205 (2005).

\bibitem{06TLB}
A. Tanguy, F. Leonforte, and J. L. Barrat, Eur. Phys. J. E \textbf{20}, 355 (2006).

\bibitem{06ML}
C. E. Maloney, and A. Lemaitre, Phys. Rev. E \textbf{74}, 016118 (2006).

\bibitem{09LP}
E. Lerner, and I, Procaccia, Phys. Rev. E \textbf{80}, 026128 (2009).

\bibitem{11RTV}
D. Rodney, A. Tanguy, and D. Vandembroucq, Modelling Simul. Mater. Sci. Eng. \textbf{19}, 08300 (2011).

\bibitem{06SLG}
G. Subhash, Q. Liu, and X-L. Gao, Int. J. Impact Engineering \textbf{32}, 1113 (2006).

\bibitem{13KTG}
A. Kara, A. Tasdemirci, and M. Guden, Materials \& Design \textbf{49}, 566 (2013).

\bibitem{13NSSMM}
 A. L. Noradila, Z. Sajuri, J. Syarif, Y. Miyashita, and Y. Mutoh, Materials Science and Engineering \textbf{46}, 012031 (2013).

\bibitem{chikkadi}
V. Chikkadi, S. Mandal, B. Nienhuis, D. Raabe, F. Varnik,
P. Schall, EPL \textbf{100}, 56001 (2012).

\bibitem{mandal}
S. Mandal, V. Chikkadi, B. Nienhuis, D. Raabe, P. Schall,
F. Varnik, Phys. Rev. E \textbf{88}, 022129 (2013).

\bibitem{varnik}
F. Varnik, S. Mandal, V. Chikkadi, D. Denisov, P. Olsson, D. Vagberg, D. Raabe, P. Schall, Phys. Rev. E \textbf{89}, 040301 (2014).

\bibitem{murali}
P. Murali, Y. W. Zhang1, and H. J. Gao, Appl. Phys. Lett. \textbf{100}, 201901 (2012).

\bibitem{cubuk}
E. D. Cubuk et.al., Science, \textbf{358}, 1033-1037 (2017).

\bibitem{nikolai}
N. V. Priezjev, Metall Mater Trans A \textbf{51}, 3713–3720 (2020).

\bibitem{Lacks1999_JCP}
D. L. Malandro, and D. J. Lacks, Journal of Chemical Physics \textbf{110}, 4593 (1999).

\bibitem{Barrat2002_PRB}
A. Tanguy, J. P. Wittmer, F. Leonforte, and J. L. Barrat, Phys. Rev. B \textbf{66}, 174205 (2002).

\bibitem{Lemaitre2004_PRL1}
C. Maloney, and A. Lema\^{i}tre, Phys. Rev. Lett. \textbf{93}, 016001 (2004).

\bibitem{Lemaitre2004_PRL2}
C. Maloney, and A. Lema\^{i}tre, Phys. Rev. Lett. \textbf{93}, 195501 (2004).

\bibitem{Lemaitre2006_PRE}
C. Maloney, and A. Lema\^{i}tre, Phys. Rev. E \textbf{74}, 016118 (2006).

\bibitem{Procaccia2009_PRE}
E. Lerner, and I. Procaccia, Phys. Rev. E \textbf{79}, 066109 (2009).

\bibitem{Kob}
W. Kob, and H. C. Andersen, Phys. Rev. E \textbf{51}, 4626 (1995).

\bibitem{Verlet}
L. Verlet, Phys. Rev. \textbf{159}, 98 (1967).

\bibitem{Berendsen}
H. J. C. Berendsen, J. P. M. Postma, W. F. van Gunsteren, A. DiNola, and J. R. Haak, J. Chem. Phys. \textbf{81}, 3684 (1984).

\bibitem{lees}
A. W. Lees and S. F. Edwards, J. Phys. C \textbf{5}, 1921 (1972).

\bibitem{jana}
R. Jana, and L. Pastewka, J. Phys. Mater. \textbf{2}, 045006 (2019).

\bibitem{chikkadi}
V. Chikkadi, and P. Schall, Phys. Rev. E \textbf{85}, 031402 (2012).

\bibitem{ding}
J. Ding, Y.Q. Cheng, and E. Ma, Appl. Phys. Lett. \textbf{101}, (2012) 121917.

\bibitem{Nikolai1}
N. V. Priezjev, Phys. Rev. E. \textbf{94}, 023004 (2016).

\bibitem{Nikolai2}
N. V. Priezjev, Phys. Rev. E. \textbf{95}, 023002 (2017).

\bibitem{vijay1}
V. Chikkadi, G. Wegdam, D. Bonn, B. Nienhuis, and P. Schall, Phys. Rev. Lett. \textbf{107}, 198303 (2011).

\end{thebibliography}
\end{document}